\documentclass[twocolumn,english,prl]{revtex4}
\usepackage[T1]{fontenc}
\usepackage[latin9]{inputenc}
\usepackage{amsmath}
\usepackage{graphicx}
\usepackage{amssymb}

\makeatletter
\@ifundefined{textcolor}{}
{%
 \definecolor{BLACK}{gray}{0}
 \definecolor{WHITE}{gray}{1}
 \definecolor{RED}{rgb}{1,0,0}
 \definecolor{GREEN}{rgb}{0,1,0}
 \definecolor{BLUE}{rgb}{0,0,1}
 \definecolor{CYAN}{cmyk}{1,0,0,0}
 \definecolor{MAGENTA}{cmyk}{0,1,0,0}
 \definecolor{YELLOW}{cmyk}{0,0,1,0}
 }

\makeatother

\makeatother

\usepackage{babel}

\makeatother

\usepackage{babel}

\makeatother

\usepackage{babel}

\makeatother

\usepackage{babel}

\makeatother

\usepackage{babel}

\makeatother

\usepackage{babel}

\begin{document}

\title{Anharmonic order-parameter oscillations and lattice coupling in strongly
driven charge-density-wave compounds: A multiple-pulse femtosecond
laser spectroscopy study}

\author{P. Kusar$^{1}$, T. Mertelj$^{1}$, V.V. Kabanov$^{1}$, J.-H. Chu$^{2}$,
I. R. Fisher$^{2}$, H. Berger$^{3}$, L. Forró$^{3}$, D. Mihailovic$^{1}$}

\affiliation{$^{1}$Complex Matter Dept., Jozef Stefan Institute, Jamova 39, Ljubljana,
SI-1000, Ljubljana, Slovenia }

\affiliation{$^{2}$Geballe Laboratory for Advanced Materials, Dept. of Applied
Physics, Stanford University, California 94305, USA}

\affiliation{$^{3}$Physics Department, EPFL CH-1015 Lausanne, Switzerland}

\date{\today}
\begin{abstract}
The anharmonic response of charge-density wave (CDW) order to strong
laser-pulse perturbations in 1$T$-TaS$_{2}$ and TbTe$_{3}$ is investigated
by means of a multiple-pump-pulse time-resolved femtosecond optical
spectroscopy. We observe remarkable anharmonic effects hitherto undetected
in the systems exhibiting collective charge ordering. The efficiency
for additional excitation of the amplitude mode by a laser pulse becomes
periodically modulated after the mode is strongly excited into a coherently
oscillating state. A similar effect is observed also for some other
phonons, where the cross-modulation at the amplitude-mode frequency
indicates anharmonic interaction of those phonons with the amplitude
mode. By analyzing the observed phenomena in the framework of time-dependent
Ginzburg-Landau theory we attribute the effects to the anharmonicity
of the mode potentials inherent to the broken symmetry state of the
CDW systems.
\end{abstract}
\maketitle
Ultrashort laser pulses are a convenient tool for coherent excitation
of phonons\cite{ZeigerVidal1992,Merlin1997} and collective electronic-lattice
modes in CDW systems\cite{DemsarBiljakovic1999,DemsarForro2002,sagarTsvetkov2007,YusupovMertelj2008}.
Due to availability of  strong laser pulses the phonons can be driven
far  from equilibrium exposing anharmonic effects. The high excitation
region has been already  investigated with single-pump-pulse\cite{HunscheWienecke1995,HaseKitajima2002,MisochkoHase2004}
and double-pump-pulse\cite{DeCampReis2001,RoeserKandyla2004,MurrayFritz2005}
sequences, mainly in elemental Bi, Sn and Te.

In charge density wave (CDW) systems the phonon mode potentials are
inherently anharmonic due to their coupling to the electron density
modulation.\cite{SchaferKabanov2010} The anharmonicity is the strongest
for the Kohn-anomaly\cite{sugai1985} modes which become the amplitude
and phase modes in the CDW state. While amplitude modes (AM) have
been extensively investigated in the near equilibrium conditions by
Raman\cite{sugai1985,LavagniniBaldini2008} and time resolved spectroscopy\cite{DemsarBiljakovic1999,DemsarForro2002,OnozakiToda2007,sagarTsvetkov2007,YusupovMertelj2008}
and in the highly driven non-equilibrium conditions, where the CDW
order is destroyed\cite{PerfettiLoukakos2006,SchmittKirchmann2008,TomeljakStadter2009,YusupovMertelj2010},
so far little attention\cite{SchmittKirchmann2008} has been paid
in the region inbetween.

Here we  report on an investigation of a new hitherto unexplored aspect
of the CDW amplitude mode behavior under strongly driven non-equilibrium
conditions in the ordered phase below the CDW photoinduced destruction
threshold. Contrary to previous standard double-pump pulse (SDPP)
high-excitation works\cite{DeCampReis2001,RoeserKandyla2004,MurrayFritz2005},
where a pair of balanced%
\footnote{The fluences of the pulses were chosen such to achieve a complete
suppression of the coherent oscillation at certain interpulse delays.%
} pump pulses was used, we introduce a novel \emph{unbalanced double-pump-pulse}
(UDPP) approach in which we use the first and the strongest pump pulse
(P$_{1}$) to excite large-amplitude coherent oscillations of the
AM and other phonon modes and then use a standard pump-probe (P$_{2}$-p$_{3}$)
pulse sequence to interrogate the system. By means of this novel approach
we are able to directly investigate the anharmonicity of the effective
AM potential, as well as detect the anharmonic coupling of the collective
bosonic mode (the AM) of the CDW to other lattice modes.

To establish generality, two layered chalcogenides which show different
types of CDW ordering and also different electronic properties, were
investigated: TbTe$_{3}$ and $1T$-TaS$_{2}$.%
\footnote{Sample growth is described in ref. {[}\onlinecite{RuChu2008}{]} for
TbTe$_{3}$ and ref. {[}\onlinecite{DardelGrioni1992}{]} for $1T$-TaS$_{2}$.%
} TbTe$_{3}$ is a two-dimensional (2D) metal which shows an unidirectional
incommensurate CDW state at the temperature used in our experiment
(15 K),\cite{DiMasiAronson1995,FangRu2007} while $1T$-TaS$_{2}$
is in a commensurate insulating CDW state at the relevant temperature
(77K)\cite{ThomsonBurk1994}. In both systems, in addition to the
AM, several new Raman modes appear in the CDW state due to Brillouin-zone
folding.\cite{sugai1985,HirataOhuchi2001,DemsarForro2002,YusupovMertelj2008,LavagniniEiter2009}

\begin{figure}[tbh]
\begin{centering}
\includegraphics[width=0.45\textwidth]{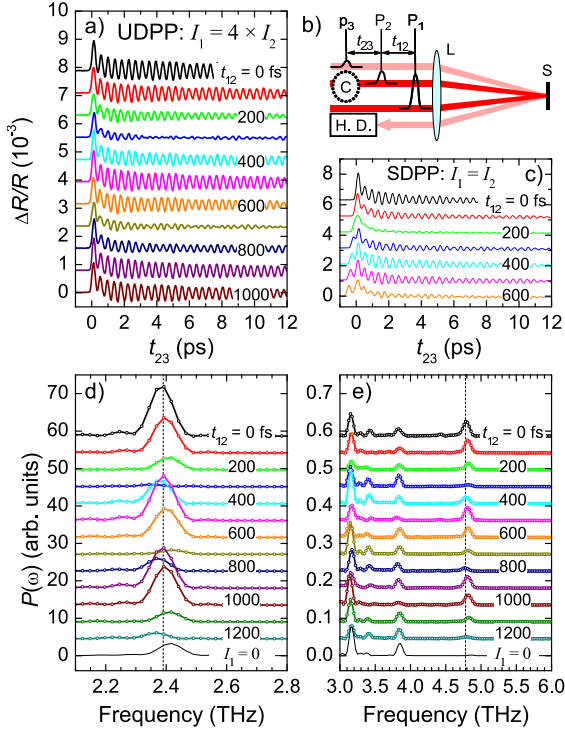} 
\par\end{centering}

\caption{(Color online) $\Delta R/R$ transients as a function of $t_{12}$
in 1$T$-TaS$_{2}$ in the UDPP configuration (a) as shown in the
schematics (b), L - lens, S - sample, C - chopper, H.D. - homodyne
detection system. For comparison $\Delta R/R$ transients in the SDPP
configuration are shown in (c). UDPP power spectra in 1$T$-TaS$_{2}$
around the fundamental frequency of the strongest mode (d) and around
the second harmonic of the fundamental frequency (e). The thin curve
in (d) and (e) is the standard single pump-pulse spectrum.}

\label{fig:fig-scans-1T-TaS2} 
\end{figure}

In our experiments the three pulse trains were derived from a 50-fs
250-kHz Ti:Al$_{2}$O$_{3}$ regenerative amplifier, with $\hbar\omega_{\mathrm{P}}=1.55$
eV photon energy. The p$_{3}$ polarization was perpendicular to P$_{1}$
and P$_{2}$, which were parallel. To study the reflectivity change
induced by the weaker P$_{2}$ pulse, $\Delta R_{2}(t)$, we eliminate
the  P$_{1}$ contribution, $\Delta R_{1}(t)$, to the total transient
reflectivity, $\Delta R(t)$, by means of the homodyne detection locked
to the modulation of the P$_{2}$ pulse train. {[}The P$_{1}$ pulse
train was unmodulated as shown in Fig \ref{fig:fig-scans-1T-TaS2}(b).{]}

In Fig. \ref{fig:fig-scans-1T-TaS2}(a) we plot the raw UDPP photoinduced
reflectivity transients, $\Delta R_{2}(t_{23})/R$, in 1$T$-TaS$_{2}$
at different delays, $t_{12}$, between the pump pulses. The intensity
of the P$_{2}$ pulse train, $I_{2}$, was set in the linear response
region while the intensity of the P$_{1}$ pulse train was 4-times
larger corresponding to $\sim$30\% of the CDW destruction treshold
fluence. For comparison the raw total photoinduced reflectivity transients
$\Delta R(t_{23})/R$, measured in SDPP configuration, are shown in
Fig. \ref{fig:fig-scans-1T-TaS2}(c). In both cases the amplitude
of the coherent oscillations periodically varies as $t_{12}$ is increased
with a clear periodic suppression of the oscillations. Note that the
suppression appears at \emph{different} $t_{12}$ for each case. While
the linear SDPP effect {[}Fig. \ref{fig:fig-scans-1T-TaS2}(c){]}
is well known\cite{DekorskyKutt1993,HaseMizoguchi1996,Mihailovic:2002p1513,OnozakiToda2007},
and is understood as an interference due to the linear superposition
of two independently excited coherent oscillations\cite{HaseMizoguchi1996,Mihailovic:2002p1513,OnozakiToda2007},
the nonlinear effects observed by UDPP are completely new and hitherto
undetected.

\begin{figure}[tbh]
\begin{centering}
\includegraphics[clip,width=0.5\textwidth]{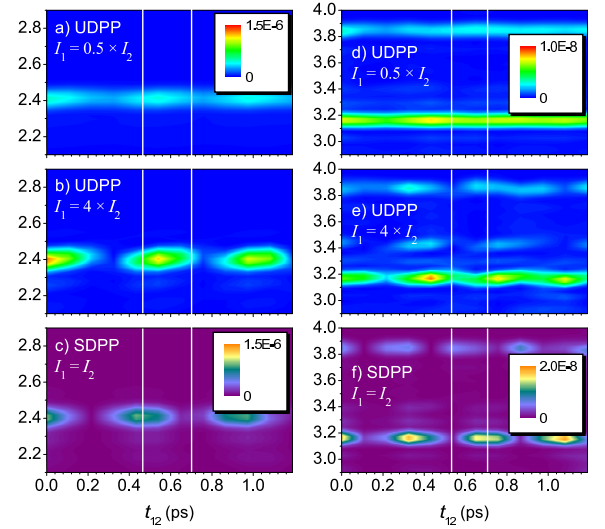} 
\par\end{centering}

\caption{(Color online) Spectra of the strongest mode (a), (b) and weaker modes
(d), (e) as functions of $t_{12}$ in the UDPP configuration at different
intensities of the P$_{1}$ pulse train in 1$T$-TaS$_{2}$. For comparison
the low-excitation SDPP-configuration spectra are shown in (c) and
(f).}

\label{fig:fig-F-dep-1TaS2-2D} 
\end{figure}

In Fig. \ref{fig:fig-scans-1T-TaS2}(d), (e) we  plot the power spectra
of the UDPP transients from Fig. \ref{fig:fig-scans-1T-TaS2}(a).
In addition to the AM mode with the frequency 2.41 THz we observe
several weaker phonon modes above 3 THz and a weak second harmonic
of the AM mode {[}see \ref{fig:fig-scans-1T-TaS2}(e){]}. The periodic
intensity modulation of the modes, which strongly increases with increasing
$I_{2}$ (see Fig. \ref{fig:fig-F-dep-1TaS2-2D}), is accompanied
with a small periodic frequency shift. The shift is absent in the
low-excitation SDPP configuration {[}see Fig. \ref{fig:fig-F-dep-1TaS2-2D}(c)
and (f){]}. The modulation amplitude and phase vary among the modes
and there is a $\sim\pi/2$ shift between the modulation phases of
the UDPP and SDPP case, similar to the observations in the high-excitation-density
SDPP experiment in Te.\cite{RoeserKandyla2004} In our case however,
the phase shift persists down to the lowest excitation density. 

\begin{figure}[tbh]
\begin{centering}
\includegraphics[width=0.35\textwidth,angle=-90]{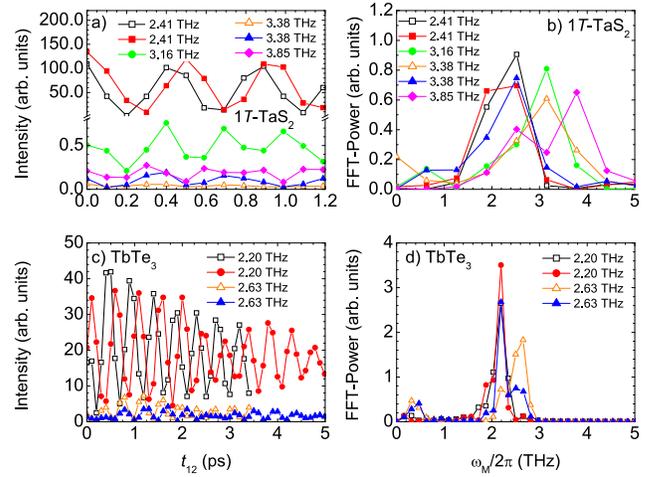} 
\par\end{centering}

\caption{(Color online) Integrated intensities of the strongest modes as functions
of $t_{12}$ in UDPP configuration at $I_{1}=4\times I_{2}$ in 1$T$-TaS$_{2}$
(a) and at $I_{1}=2.4\times I_{2}$ in TbTe$_{3}$ (c). Normalized
 power spectra of the traces in the left panels (b) and (d). Open
symbols correspond to the low-intensity SDPP response. Note the difference
in the modulation frequency for the 3.38-THz mode in $1T$-TaS$_{2}$
(b) and 2.63-THz mode in TbTe$_{3}$ (d) between the UDPP (full triangles)
and SDPP (open triangles) cases.}

\label{fig:fig-IvsT12-both} 
\end{figure}

The periodic intensity modulation and the phase shift are observed
also in TbTe$_{3}$, where in addition a beating in the $t_{12}$
dependence of the modulation amplitudes is observed (see Fig. \ref{fig:fig-IvsT12-both}(c)).%

The modulation frequency of each mode {[}see Fig. \ref{fig:fig-IvsT12-both}
(b) and (d){]} is correlated to the respective mode eigenfrequency.
Surprisingly, in the UDPP configuration there is also a clear \emph{cross-modulation}
of the 3.38-THz and 3.85-THz mode intensities with the 2.41-THz AM
frequency and the 2.63-THz mode intensity with the 2.20-THz AM frequency
in 1$T$-TaS$_{2}$ and TbTe$_{3}$, respectively.

To understand the observed phenomena we start with the simplest Ginzburg-Landau
expansion of the free energy:

\begin{equation}
F=F_{0}+\left(\frac{T}{T_{\mathrm{c}}}-1\right)|A|^{2}+\frac{1}{2}|A|^{4}+g(t)|A|^{2},\label{eq:GL}\end{equation}
in terms of the normalized complex order parameter, $A$. $T_{c}$
is the critical temperature and $g(t)$ represents the external laser
excitation. Due to the symmetry $g(t)$ can only couple to $|A|^{2}$.
To describe the dynamics we introduce the $T=0$ AM frequency, $\omega_{0}$,
and the dimensionless damping, $\gamma=\Delta\omega_{0}/\omega_{0}$,
and obtain using (\ref{eq:GL}): \begin{equation}
\frac{2}{\omega_{0}^{2}}\frac{\partial^{2}}{\partial t^{2}}A+\frac{4\gamma}{\omega_{0}}\frac{\partial}{\partial t}A+\left(\frac{T}{T_{\mathrm{c}}}-1\right)A+|A|^{2}A=-g(t)A.\label{eq:dynamic}\end{equation}
Since $g(t)A$ can not excite phase fluctuations%
\footnote{In the equilibrium, $A$ can be always be set real by a proper choice
of the phase.%
}, only the AM needs to be considered. The dielectric constant depends
in the lowest order on $|A|^{2}$:\cite{GinzburgLevanyuk1980}\begin{equation}
\epsilon=\epsilon_{0}+c_{1}|A|^{2},\label{eq:eps}\end{equation}
so the reflectivity change in the UDPP configuration is written as:\begin{equation}
\Delta R_{2}(t)=\Delta R(t)-\Delta R_{1}(t)\propto|A(t)|^{2}-|A_{1}(t)|^{2},\label{eq:dr2}\end{equation}
where $A_{1}(t)$ is the solution of equation (\ref{eq:dynamic})
with excitation from the P$_{1}$ pulse only and $A(t)$ the solution
with the excitation from both pump pulses.

There are two terms in (\ref{eq:dynamic}) that are of interest: $g(t)A$
and $|A|^{2}A$. The term $g(t)A$ leads to a periodic modulation
of the coupling of $A$ to the P$_{2}$ pulse when the oscillation
amplitude after the P$_{1}$ pulse is large. This effect however does
not explain the main features of our observations: (i) if the oscillation
amplitude is large, a significant AM-overtone intensity is expected
due to $|A|^{2}$ in (\ref{eq:eps}), which is not observed in the
experiment, (ii) because the laser excitation initially drives $A$
towards zero%
\footnote{$g(t)$ couples to $A$ as $T$ so a positive $g(t)$ decreases $A$.%
} an intermediate amplitude of the oscillations at $t_{12}=0$ followed
by a minimum at $\omega_{0}t_{12}\simeq\pi/2$, where $A(t)$ is closest
to 0, is expected. Instead, the first minimum is observed around $\omega_{0}t_{12}\simeq3\pi/2$
and the amplitude is the largest at $t_{12}=0$ in 1$T$-TaS$_{2}$
and at $\omega_{0}t_{12}\simeq\pi/2$ in TbTe$_{3}$. The anharmonic
term $|A|^{2}A$ in (\ref{eq:dynamic}) remains therefore the only
possible origin of the observed behavior.

\begin{figure}[tbh]
\begin{centering}
\includegraphics[angle=-90,width=0.45\textwidth]{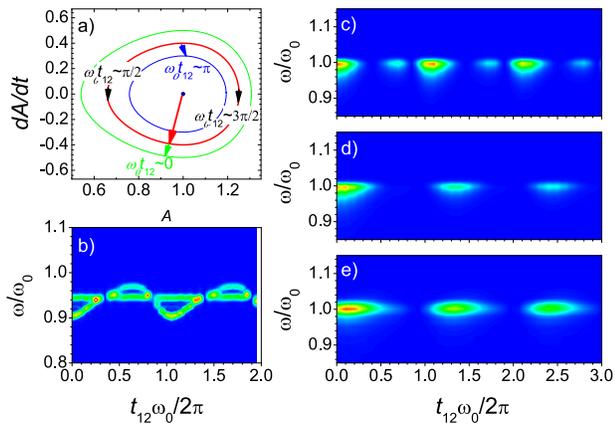} 
\par\end{centering}

\caption{(Color online) Orbits of eq. (\ref{eq:dynamic}) in the phase space
in the absence of damping ($\gamma=0$) (a). The long arrow represents
the initial excitation by the P$_{1}$ pulse. The short arrows represent
the additional excitation by the P$_{2}$ pulse which transfers the
system to different orbits depending on $t_{12}$. Simulated power
spectrum as a function of $t_{12}$ in the absence of damping (b),
with damping ($\gamma=0.01$) and short excitation pulses, $\omega_{0}\tau_{g}=0.2\pi$
(c), with damping and long excitation pulses $\omega_{0}\tau_{g}=2\pi$
(d), with damping, long excitation pulses and finite laser penetration
depth, $\lambda/\xi$=16, (e). }

\label{fig:fig-orbits} 
\end{figure}

The solutions of equation (\ref{eq:dynamic}) with $\gamma=0$ can
be represented as closed periodic orbits in the phase space (see Fig.
\ref{fig:fig-orbits} (a)). Due to the anharmonicity the frequency
of the orbit decreases with the oscillation amplitude. The P$_{2}$
pulse transfers the system from the initial orbit, set by the P$_{1}$
pulse, to a final orbit with a different frequency which depends on
$t_{12}$. This results in a beating of the $\Delta R_{2}(t)$ oscillations
due to an interference of the second and the first term in the right
hand side of (\ref{eq:dr2}) corresponding to the initial and final
orbits, respectively. The frequency of the final orbit periodically
oscillates with increasing $t_{12}$ {[}see Fig. \ref{fig:fig-orbits}
(b){]}. Within a single initial orbit period there exist two delays,
$\omega_{0}t_{12}\sim\phi_{0}+\pi/2$ and $\sim\phi_{0}+3\pi/2$,
at which the only effect of the P$_{2}$ pulse is a phase shift within
the initial orbit. In vicinity of these delays the beating is very
slow resulting in a slow increase of the $\Delta R_{2}(t)$ oscillations
amplitude with $t$. This slow rise is suppressed when $\gamma$ is
finite since the oscillations die out before a significant phase shift
between the orbits builds up and the $\Delta R_{2}(t)$ oscillations
amplitude remains small. As a result a periodic modulation of the
$\Delta R_{2}(t)$ oscillations amplitude is observed in the simulations
with finite $\gamma$ {[}see Fig. \ref{fig:fig-orbits} (c){]}. In
addition to the more intensive maximum, reminiscent to the experimental
observations, an additional weak maximum is observed within a single
period corresponding to $t_{12}$ with the the larger final-orbit
frequency {[}see Fig. \ref{fig:fig-orbits} (b){]}. The weak maximum
is strongly sensitive to the characteristic timescale of the external
laser perturbation, $\tau_{g}$,%
\footnote{In simulations $g(t)=g_{1}f(t)+g_{2}f(t-t_{12})$ with $f(t)=\exp(-t/\tau_{g})/[1+\exp(-10t/\tau_{g})]$
was used to represent P$_{1}$ and P$_{2}$.%
} and is completely suppressed by setting $\tau_{g}=2\pi/\omega_{0}$
as shown in Fig. \ref{fig:fig-orbits} (d).

To improve agreement of the model with the experiment we extended
the Ginzburg-Landau expansion (\ref{eq:GL}) with the gradient term,
$\xi^{2}|\partial A/\partial z|^{2}$,\cite{YusupovMertelj2010} to
allow for space variations of the order parameter perpendicular to
the sample surface due to the finite penetration depth, $\lambda$,
of the laser pulses. The result of a simulation with the inhomogeneous
order parameter {[}Fig. \ref{fig:fig-orbits} (e){]} better reproduces
the main experimental features for the AM. Similarly to the homogeneous
case the weak maximum, which is not observed in the experiment, is
absent only for long enough $\omega_{0}\tau_{g}\sim2\pi$. In our
experiment $\omega_{0}\tau_{l}\sim0.2\pi$, where $\tau_{l}$ is the
length of the laser pulses, indicating that the coherent oscillations
are not excited by the impulsive stimulated-Raman-scattering mechanism\cite{Merlin1997},
but rather by the displacive one\cite{ZeigerVidal1992}, which can
lead to $\tau_{g}>\tau_{l}$.

Counter intuitively, the experimentally observed small periodic frequency
modulation of the AM is not reproduced in any simulation with a finite
damping so the origin of the effect is beyond the present model.%

Next we turn to analysis of the weaker modes with displacements denoted
$Q_{i}$. Similarly to the AM we can exclude the direct driving terms,
$|Q_{i}|^{2}g(t)$ and $\Re(Q_{i}A^{*})g(t)$, as the sources of the
modulation due to the relatively small displacements. However, in
addition to the AM also other modes are coupled to the CDW charge
modulation. As a result, mode displacements $Q_{i}$ become mixed
with the AM displacement%
\footnote{The terms $\Re(Q_{i}A^{*})$ must be added to the total free energy.%
} $A$ and the effective mixed-mode potentials become anharmonic. This
leads to the \emph{cross-modulation} of mode intensities with the
AM frequency in addition to the self-modulation at the particular
mode eigenfrequency. The relative amounts of the modulation at both
frequencies depend on the couplings to the AM and to the external
laser perturbation. If after the P$_{1}$ pulse the amplitude of a
particular mode is small the main contribution to the modulation is
due to the cross-modulation at the AM frequency, as is the case for
the 3.38-THz mode in 1$T$-TaS$_{2}$ and the 2.63-THz mode in TbTe$_{3}$.
If, on the other hand, the amplitude is large enough a significant
self-modulation is expected as in the case of the 3.85-THz in 1$T$-TaS$_{2}$.

In conclusion, by means of a novel \emph{unbalanced double-pump-pulse}
time resolved optical spectroscopy, we were able to detect the inherent
broken-symmetry-state anharmonicity of the amplitude-mode effective
potential in two distinct CDW systems. We also found a clear evidence
of the anharmonic mixing of certain phonon modes with the amplitude
mode originating from their mutual coupling to the electronic-density
CDW modulation. We showed that the observed effects can be described
in the framework of time dependent Ginzburg-Landau theory. 
\begin{acknowledgments}
This work has been supported by Slovenian Research Agency and the
CENN Nanocenter.
\end{acknowledgments}
\bibliography{biblio}

\end{document}